# Strain-tunable band gap of a monolayer graphene analogue of ZnS monolayer

Harihar Behera and Gautam Mukhopadhyay[*]

*Department of Physics, Indian Institute of Technology Bombay, Powai, Mumbai-400076, India*

[*]*Corresponding author, E-mail: gmukh@phy.iitb.ac.in*

**Abstract:** Using first-principles full-potential density functional calculations, we predict that mechanically tunable band-gap is realizable in ZnS monolayer in graphene-like honeycomb structure by application of in-plane homogeneous biaxial strain. A transition point from direct-to-indirect gap-phase is predicted to exist for biaxial tensile strain lying in the interval (2.645%, 3.171%). In the two gap-phases, the band gap decreases with increasing strain and varies linearly with strain.

## Introduction

The study of two dimensional (2D) materials is an emerging field of research inspired by the recent phenomenal growth in the research on graphene (a one atom-thick 2D nanocrystal of carbon atoms in a hexagonal lattice) promising many novel applications [1]. A number of 2D/quasi-2D nanocrystals of BN [1], $MoS_2$ [1], $MoSe_2$ [1], $Bi_2Te_3$ [1], Si [2], ZnO [3] have been synthesized. Using density functional theory (DFT) calculations, recently Freeman et al [4] predicted that when the layer number of (0001)-oriented wurtzite (WZ) materials (e.g., AlN, BeO, GaN, SiC, ZnO and ZnS) is small, the WZ structures transform into a new form of stable hexagonal BN-like structure. This prediction has recently been confirmed in respect of ZnO [3], whose stability is attributed to the strong in-plane $sp^2$ hybridized bonds between Zn and O atoms. The atomic structure and chemical properties of ZnS are comparable to that of ZnO [5, 6]. ZnS is a direct band gap ($E_g \sim 3.72$ eV in normal cubic zinc blend phase and $E_g \sim 3.77$ eV in hexagonal wurtzite (WZ) phase) semiconductor which promise a variety of novel applications as light-emitting diodes (LEDs), electroluminescence, flat panel displays, infrared windows, photonics, lasers, sensors, and biodevices etc. [5, 6]. Further, due to its larger band gap than that of ZnO ($E_g \sim 3.4$ eV), ZnS is more suitable for visible-blind ultraviolet (UV)-light sensors/photodetectors than ZnO [5]. Although ZnS nanostructures in the form of nanoclusters, nanosheets, nanowires, nanotubes and nanobelts are currently being intensely studied both experimentally and theoretically [5, 6], the study of a monolayer graphene (MLG) analogue of ZnS (ML-ZnS) is scanty [7]. Here, we report our DFT calculations on ML-ZnS without and with application of in-plane homogeneous biaxial strain.

## Calculation methods

The calculations have been performed by using the DFT based full-potential (linearized) augmented plane wave plus local orbital (FP-(L)APW+lo) method [8] as implemented in the elk-code [9]. We have used the Perdew-Zunger variant of LDA [10], the accuracy of which has been successfully tested in our previous studies [11-13]. For plane wave expansion in the interstitial region, we have used $|G+k|_{max} \times R_{mt} = 8$ where $R_{mt}$ is the smallest muffin-tin radius, for deciding the plane wave cut-off. The Monkhorst-Pack [14] **k**-point grid size of 20×20×1 was used for structural and of 30×30×1 for band structure and density of states (DOS) calculations. The total energy was converged within 2μeV/atom. We simulate the 2D-hexagonal structure of ML-ZnS as a 3D-hexagonal supercell with a large value of *c*-parameter (= |**c**| = 40 a.u.). The application of homogeneous in-plane biaxial strain δ < 6% was simulated by varying the in-plane lattice parameter *a* (=|**a**| = |**b**|); δ = (*a* – $a_0$)/$a_0$, where $a_0$ is the ground state in-plane lattice constant. Figure 1 (a) depicts the top-down view of ML-ZnS in planar graphene-like honeycomb structure.

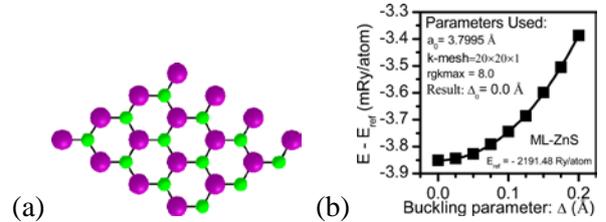

(a)  (b)

**Figure 1.** (a) Top-down view of flat ML-ZnS in ball-stick model. (b) Probing buckling in ML-ZnS using the principle of minimum energy for the stable structure; in buckled ML-ZnS, Zn and S atoms take positions in two different parallel planes, buckling parameter 'Δ' is the perpendicular distance between those parallel planes and Δ = 0.0 Å for a flat ML-ZnS.

## Results and discussions

For an assumed flat ML-ZnS, our calculated LDA value of $a_0$ = 3.7995 Å corresponding to the Zn-S bond length $d_{Zn-S} = a_0/\sqrt{3}$ = 2.194 Å is in agreement with theoretical GGA value of $d_{Zn-S}$ = 2.246 Å for ZnS flat single sheet [7] considering the fact that GGA usually overestimates the lattice constant. Our assumption on the flat ML-ZnS structure was tested as correct by the calculated results depicted Figure 1(b).

The band structure and total DOS (TDOS) plot (Figure 2) show that ML-ZnO is a direct band gap ($E_g$ = 2.622 eV, LDA value) semiconductor with both valence band maximum (VBM) and conduction band minimum (CBM) located at the Γ point of the hexagonal Brillouin Zone (BZ). However, the actual band gap is expected to be larger since LDA underestimates the gap. Our calculated LDA band gap of 2.622 eV is larger than the reported GGA value of $E_g$ = 2.07 eV for ZnS single sheet [7] and the GGA value of $E_g$ = 1.68 eV for ML-ZnO [15, 16].

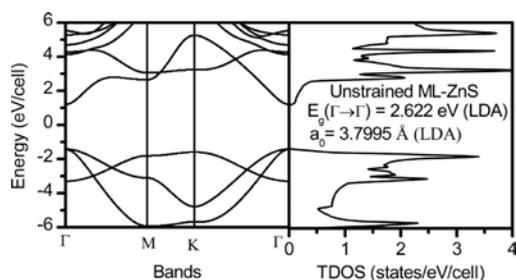

**Figure 2.** Bands and total DOS (TDOS) of unstrained ML-ZnS within LDA. Fermi energy level is at 0 eV.

The nature of variations of our calculated values of $E_g$ with δ for flat ML-ZnS is depicted in Figure 3. As seen in Figure 3, a transition point from direct-to-indirect gap-phase exists for 'a' value lying in the interval (3.90 Å, 3.92 Å) and in the two gap-phases, the variation of $E_g$ with δ is approximately linear. In the indirect gap-phase, VBM is at the K and CBM is at the Γ point of the BZ.

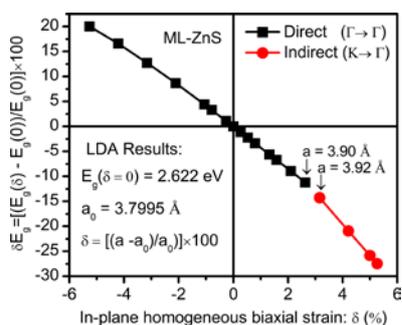

**Figure 3.** Relative variation of $E_g$ of flat ML-ZnS with in-plane homogeneous biaxial strain δ = $(a - a_0)/a_0$.

## Conclusions

While corroborating the previous prediction of a possible planar graphene-like stable structure of ML-ZnS, we predict a strain-tunable band gap of flat ML-ZnS. We hope, with the advancement of fabrication techniques, these predictions are testable in near future for potential applications in a variety of novel nano-devices such as strain sensors, mechatronics, nano-electromechanical systems (NEMS) and nano-optomechanical systems (NOMS) etc.

## References

1. Neto A. H. C., and Novoselov K. New directions in science and technology: two-dimensional crystals, Rep. Prog. Phys. **74** (2011) 1-9.
2. Kara A., Enriquez H., et al. A review on silicene – New candidate for electronics, Surf. Sci. Rep. **67** (2012) 1-18.
3. Tusche C., et al. Observation of depolarized ZnO(0001) monolayers: Formation of unreconstructed planar sheets, Phys. Rev. Lett. **99** (2007) 026102.
4. Freeman C.L., Claeyssens F., et al. Graphitic nanofilms as precursors to wurtzite films: Theory, Phys. Rev. Lett. 96 (2006) 066102.
5. Fang X., Zhai T., et al. ZnS nanostructures: From synthesis to applications, Prog. Mater. Sci., **56** (2011) 175.
6. Fang X., Bando Y., et al. ZnS Nanostructures: ultraviolet-light emitters, lasers, and sensors, Crit. Rev. Solid State Mat. Sci., **34** (2009) 190-223.
7 . Krainara N., et al. Structural and electronic bistability in ZnS single sheets and single-walled nanotubes, Phys. Rev. B **83** (2011) 233305.
8. Sjöstedt E., et al. An alternative way of linearizing the augmented plane-wave method, Solid State Commun. **114**, (200) 15-20.
9. Elk is an open source code: http://elk.sourceforge.net/
10. Perdew P., Zunger A. Self-interation correction to density-functional approximations for many electron systems, Phys. Rev. B **23** (1981) 5048.
11. Behera H., Mukhopadhyay G. Structural and electronic properties of graphene and silicene, AIP Conf. Proc.**1313** (2010) 152-155; arXiv:1111.1282
12. Behera H., Mukhopadhyay G. First-Principles Study of Structural and Electronic Properties of Germanene, AIP Conf. Proc.**1349** (2011) 823-824; arXiv:1111.6333
13. Behera H., Mukhopadhyay G. Strain-tunable band gap in graphene/h-BN hetero-bilayer, J. Phys. Chem. Solids **73** (2012) 818-821; arXiv:1204.2030
14. Monkhorst H.J., Pack J.D. Special points for Brillouin-zone integrations, Phys. Rev. B **13** (1976) 5188-5192.
15. Topsakal M., et al. First-principles study of Zinc Oxide honeycomb structures, Phys. Rev. B **80** (2009) 235119.
16. Behera H., Mukhopadhyay G. Strain-tunable direct band gap of ZnO monolayer in graphene-like honeycomb structure, arXiv:1111.6322